\documentclass[ reprint, amsmath,amssymb, aps, onecolumn,11pt,tightenlines]{revtex4-2}

\usepackage{caption}
\usepackage{tabularx}
\usepackage{amssymb}
\usepackage{mathrsfs}  
\usepackage{amsmath}
\usepackage{graphicx}
\usepackage{dcolumn}
\usepackage{bm}
\usepackage{wrapfig}

\def\del#1{}

\def\outline#1{\textcolor{Gray}{}}

\begin{document}

\preprint{APS/123-QED}

\title{Green's Function Integral method for Pressure Reconstruction from Measured Pressure Gradient and the Interpretation of Omnidirectional Integration}

\author{Qi Wang}
\author{Xiaofeng Liu}%
 \email{xiaofeng.liu@sdsu.edu }
\affiliation{%
 Department of Aerospace Engineering, San Diego State University.}%





\begin{abstract}
Accurately and efficiently measuring the pressure field is of paramount importance in many fluid mechanics applications. The pressure gradient field of a fluid flow can be determined from the balance of the momentum equation based on the particle image velocimetry (PIV) measurement of the flow kinematics, which renders the experimental evaluation of the material acceleration and the viscous stress terms possible. We present a novel method of reconstructing the pressure fields from error-embedded pressure gradient measurement data. This method utilized Green's function of the Laplacian operator as the convolution kernel that relates pressure to pressure gradient.
A compatibility condition on the boundary offers equations to solve for the boundary pressure.
This Green's function integral (GFI) method has a deep mathematical connection with the state-of-the-art omnidirectional integration (ODI). As mathematically equivalent to ODI in the limit of an infinite number of line integral paths, GFI spares the necessity of line integration along zigzag integral paths, rendering generalized implementation schemes for both two and three-dimensional problems with arbitrary geometry while bringing in improved computational efficiency. 
In the current work, we apply GFI to pressure reconstruction of simple canonical and isotropic turbulence flows embedded with error in two-dimensional and three-dimensional domains, respectively.
Uncertainty quantification is performed by eigenanalysis of the GFI operator in a square domain.
The accuracy and computational efficiency of GFI are evaluated and compared with ODI.
\end{abstract}

\maketitle


\section{Pressure field estimation from PIV}
Pressure is of paramount importance for characterization and modeling of a variety of flow phenomena and engineering applications, e.g., unsteady and turbulent flows \citep{blake2017mechanics}, pressure-related turbulence modeling  \citep{girimaji2000pressure, liu2018pressure, pope2000turbulent}, cavitation \citep{arndt2002cavitation, brennen2014cavitation}, acoustics \citep{haigermoser2009application, liu2011time} and lift and form drag generation over a moving body in fluid \citep{anderson2011ebook}, etc. Accompanying the advancement in Particle Image Velocimetry (PIV) theory, measurement techniques and data post-processing algorithms, it has been shown over the past two decades that the instantaneous spatial pressure distribution in a turbulent flow field can be measured non-intrusively using PIV, as demonstrated through a variety of experimental efforts, e.g., \cite{liu2006instantaneous, violato2011lagrangian, moore2011two, van2013piv, van2017comparative}, to name a few.  


As introduced in \cite{liu2003measurements} and \cite{liu2006instantaneous}, the non-intrusive pressure measurement technique stems from the Navier-Stokes equation,  
\begin{equation}
        \nabla p = - \rho\frac{D \boldsymbol{u}}{Dt} + \mu \nabla^2 \boldsymbol{u} 
\end{equation}
from which, the pressure gradient can be determined non-intrusively by measuring the material acceleration and the viscous terms using PIV. Once the pressure gradient is obtained, further integration of the measured pressure gradient using appropriate pressure reconstruction algorithms can then lead to the reconstructed pressure distribution over the measurement domain. Thus, a complete experimental campaign for obtaining pressure distribution encompasses both an initial stage of kinematic quantity measurement for obtaining the pressure gradient and a subsequent stage of data post-processing for obtaining the reconstructed pressure from the measured pressure gradient. 

For high Reynolds number flows and for regions away from solid walls, the material acceleration is the dominant contributor to the pressure gradient \citep{liu2006instantaneous}.  Theoretically, the material acceleration can be obtained using either the Eulerian approach \citep{de2012instantaneous, violato2011lagrangian, gurka1999computation}, where the material acceleration is calculated indirectly through the Eulerian expansion of the material acceleration into the unsteady and convection terms, or the Lagrangian approach where the material acceleration is calculated directly by tracing imaginary fluid particles along their trajectories based on the so-called pseudo tracking method \citep{liu2003measurements, de2012instantaneous, violato2011lagrangian, jensen2003experimental}.  Through investigations of error propagation from velocity to acceleration, \cite{jensen2004optimization, violato2011lagrangian, de2012instantaneous, van2013piv, wang2016irrotation} and \cite{van2017comparative} demonstrate that for advection-dominated flows, the Lagrangian approach shows consistently
less sensitive to noise than the Eulerian approach. Therefore the Lagrangian approach is the method of choice for the material acceleration measurement using PIV. 

Even with thoughtful consideration and careful implementation of the PIV measurement for the flow kinematics, errors associated with the measurement data are still inevitable. These errors impose great challenges to pressure reconstruction from the measured pressure gradient. This situation, as shown in \cite{liu2020error}, is especially acute when one tries to solve the Poisson equation to obtain the reconstructed pressure from the error-embedded pressure gradient using Neumann boundary condition, which in many cases is the only available condition for a Poisson solver approach.

To overcome this difficulty, \cite{liu2003measurements, liu2006instantaneous} introduced the original Omni-Directional
Integration (ODI) algorithm to minimize the influence of the pressure gradient measurement error on the final reconstructed pressure result. Subsequently, \cite{liu2016instantaneous} improved the algorithm to the Rotating Parallel Ray Omni-Directional Integration, a state-of-the-art version of its kind. \cite{liu2020error} and \cite{liu2021pressure} further demonstrated that for both simply-connected  and multiply-connected flow domains, ODI methods outperform the conventional Poisson equation approach in pressure reconstruction accuracy, with the parallel ray
being the algorithm with the best performance in accuracy among all ODI methods. Essentially, as shown in \cite{liu2020error} based on theoretical analysis, the ODI algorithms provide an effective boundary pressure error reduction mechanism through iteration, which upon final convergence, effectively renders Dirichlet boundary conditions to ensure accurate pressure reconstruction over the entire measurement domain.  

However, although the computation cost of using a 2D version of the parallel ray omnidirectional integration is acceptable (less than a minute) for calculating a planar measurement domain with $250 \times 250$ nodal points, the computation cost for a 3D domain with a typical tomo-PIV vector space of $100\times50\times50$ grid is daunting. Therefore, the 3D implementation of the parallel ray omnidirectional integration algorithm \citep{wang2019gpu} has to rely on GPU-based computation to accelerate the computation speed and cut the computation time. Motivated by this shortcoming, in this paper we introduce a novel method to reconstruct the
pressure field from error-embedded pressure gradients. This method utilizes Green’s function of the Laplacian operator and relates pressure to pressure gradient through a convolution kernel. The compatibility condition offers equations to solve for the boundary pressure. In this paper, we demonstrate that Green’s function integral (GFI) method is mathematically equivalent to omnidirectional integration (ODI) in the limit of an infinite number of ODI line integral paths. Yet GFI spares the necessity of line integration along zigzag integral paths in ODI, rendering general schemes easier to be implemented and more affordable for both two and three-dimensional problems with arbitrary inner and outer boundary geometry shapes. 

The paper is organized in the following fashion. In \S\ref{Sec:numericalmethods}, we discuss the theory and formulation of Green’s function integral method, as well as the mathematical relationship between the GFI and the ODI methods. In \S\ref{Sec:impulse}, we analyze the impulse response for GFI and ODI to demonstrate the mathematical equivalence. \S\ref{Sec:2D} applies GFI to pressure reconstruction of simple canonical and isotropic turbulence flows embedded with error in a two-dimensional domain and shows that the accuracy of GFI is equivalent to that of ODI, yet GFI has much more significantly improved computational efficiency in comparison with ODI. In \S\ref{Sec:2DUQ} we perform eigenanalysis for the GFI operator in two-dimensional setups to demonstrate the denoising mechanism of GFI. The framework is further briefly extended to a three-dimensional problem with complex geometry in \S\ref{Sec:3D}.

\section{Green's function integral method}
\label{Sec:numericalmethods}
\subsection{Field inversion from pressure gradient to pressure using Green's function}
The pressure at a point $\boldsymbol{x}^\prime$ inside domain $\Omega$ can be evaluated from the integration between the pressure field and a Dirac delta function at that point, i.e. $\displaystyle p(\boldsymbol{x}^\prime) = \int_\Omega p(\boldsymbol{x}) \delta(\boldsymbol{x}-\boldsymbol{x}^\prime) d\boldsymbol{x}$.
Suppose that the Green's function for the Laplacian operator is $G(\boldsymbol{x},\boldsymbol{x}^\prime)$, satisfying 
\begin{equation}
    \nabla^2 G(\boldsymbol{x}, \boldsymbol{x}^\prime) = \delta(\boldsymbol{x}^\prime-\boldsymbol{x}),
\end{equation}
where $\nabla^2$ represents the Laplacian with respect to the variable $\boldsymbol{x}$.
It can be proven that in two-dimensional space, $\displaystyle G(\boldsymbol{x},\boldsymbol{x}^\prime) = \frac{1}{2\pi}\ln |\boldsymbol{x}-\boldsymbol{x}^\prime|$, while in three-dimensional space $\displaystyle G(\boldsymbol{x},\boldsymbol{x}^\prime) = -\frac{1}{4\pi |\boldsymbol{x}-\boldsymbol{x}^\prime|}$ \citep{prosperetti2011advanced,thijssen1998mathematical}.
For convenience, we denote $r = |\boldsymbol{x} - \boldsymbol{x}^\prime| $ and the Green's function can be written as a function of $r$.

Applying the divergence theorem to the vector field $p \boldsymbol{\nabla} G$ yields,
\begin{equation}
\begin{aligned}
    \oint_{\partial \Omega}p \boldsymbol{\nabla} G \cdot d\boldsymbol{S} 
     = \int_{\Omega} \boldsymbol{\nabla} \cdot \left(p \boldsymbol{\nabla} G\right) dV 
     =  \int_{\Omega}  \boldsymbol{\nabla} p \cdot\boldsymbol{\nabla} G \; dV +  \int_{\Omega} p \nabla^{ 2} G \; dV.
\end{aligned}
\end{equation}
This above identity is also coined as the first Green's identity. 
The last term on the right-hand side is,
\begin{equation}
    \int_{\Omega} p \nabla^{ 2} G \; dV = \int_{\Omega} p\delta(\boldsymbol{x}-\boldsymbol{x}^\prime) dV = p(\boldsymbol{x}^\prime).
\end{equation}
Combining the above two equations, the pressure at $\boldsymbol{x}^\prime$ inside the domain $\Omega$ can be related to the pressure gradient at other locations $\boldsymbol{x}$, specifically,
\begin{equation}
\begin{aligned}
    p(\boldsymbol{x}^\prime) 
    &= -\int_\Omega \boldsymbol{\nabla} p \cdot \boldsymbol{\nabla} G \; dV
     + \oint_{\partial \Omega} p \boldsymbol{\nabla} G \cdot  d\boldsymbol{S}.
     \label{eqn:GFI}
\end{aligned}
\end{equation}
The determination of pressure can be split into two parts: the convolution between the pressure gradient and the kernel $\boldsymbol{\nabla} G$, and the contribution from the boundary pressure.
The convolution kernel $\boldsymbol{\nabla} G$ is the same as the velocity field for a potential source flow with unit intensity, which implies the influence for a perturbation in the pressure gradient is similar to a potential source flow.

\subsection{Compatibility condition on the boundary}
Equations for pressure on the boundary $\partial \Omega$ would involve the Dirac delta function at the boundary of the domain.
Normally, such a delta function is ill-defined and its property is problem-dependent.
In the current setup, we regard the delta function as a heuristic ``radial" function that only depends on $r$.
Therefore, it can be defined that $\displaystyle \int_\Omega p(\boldsymbol{x}) \delta(\boldsymbol{x}-\boldsymbol{x}^\prime) d\boldsymbol{x} =\frac 12 p(\boldsymbol{x}^\prime)$ 
for $\boldsymbol{x}$ on the boundary $\partial \Omega$, and $\partial \Omega$ is smooth at $\boldsymbol{x}^\prime$.
\begin{equation}
\begin{aligned}
    p(\boldsymbol{x}^\prime) = -2\int_\Omega \boldsymbol{\nabla} p \cdot \boldsymbol{\nabla} G  \;dV  + 2\oint_{\partial \Omega} p  \boldsymbol{\nabla} G\cdot  d\boldsymbol{S}, \quad \boldsymbol{x}^\prime \in \partial \Omega.
     \label{eqn:GFI_bc}
\end{aligned}
\end{equation}
Therefore, if we assume $N_b$ grid points on the boundary, the above equation would result in $N_b$ linear equations coupling all pressure on the boundary together, from which we can solve for the pressure field on the boundary. 
Notice that adding a constant to the pressure field does not change the pressure gradient field, representing one redundant equation. In practice, we also replace one equation with the averaged boundary pressure assuming zero value in order to invert the system.

\subsection{Numerical Discretization}
Based on equation \ref{eqn:GFI} and \ref{eqn:GFI_bc}, an algorithm to evaluate the pressure distribution both inside and on the boundary of the domain can be formulated. Throughout this study, we adopted a finite element type of approach that can be applied to both structured and unstructured mesh with any shape.
\begin{figure}
    \centering
    \includegraphics[width = \textwidth]{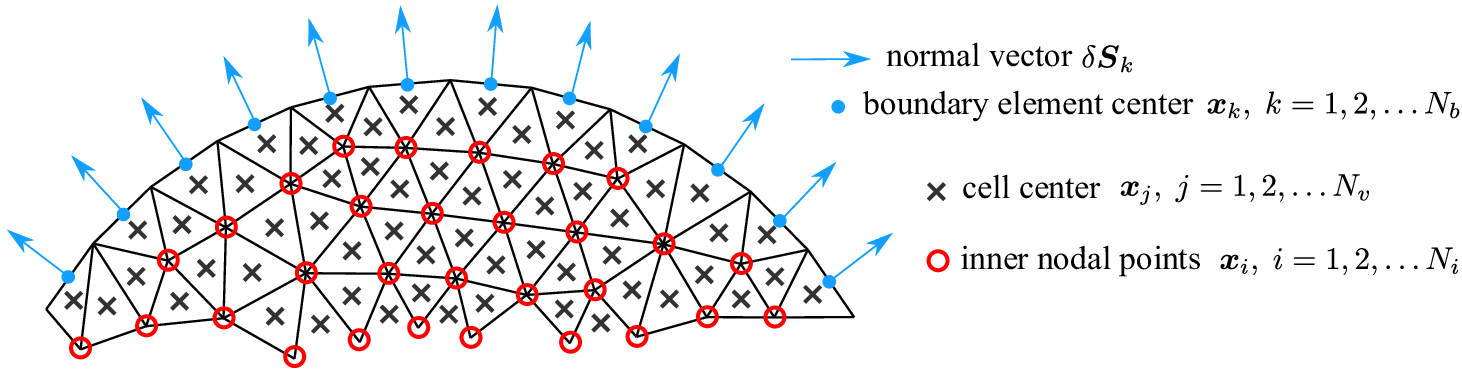}
    \caption{Schematic of the mesh}
    \label{fig:mesh}
\end{figure}

We divide the domain $\Omega$ into $N_v$ elements with volume $\delta V_j, \; j= 1,2,\ldots, N_v$. As shown in the schematic (Figure \ref{fig:mesh}), the pressure gradients, i.e. $\left(\boldsymbol{\nabla}p\right)_j, \; j=1,2,\ldots, N_v$, are given at the cell centers while the pressure fields $p$ is to be determined on the inner nodal points (marked by red circles in the schematic). 
Suppose that the number of boundary elements is $N_b$, the boundary pressure $p_{b,k},\; k= 1,2,\ldots, N_b$ are located at the center of the boundary segment, with normal vector $\delta \boldsymbol{S}_k$ pointing out of the domain and perpendicular to the surface element. 
Throughout the paper, we use indices $i,j$, and $k$ for inner nodal points, cell centers, and boundary element centers, respectively.
For a two-dimensional setup, the magnitude of $\delta \boldsymbol{S}_k$ is the length of the line segment of the boundary element, while in a three-dimensional setup, the magnitude is the area of the surface elements. 

For two-dimensional ($D=2$) and three-dimensional ($D=3$) problems, the unified expression for the gradient of the Green's function is $\displaystyle \boldsymbol{\nabla} G = \frac 1 {2(D-1)\pi }\frac{\boldsymbol{r}}{r^D}$, and the equation \ref{eqn:GFI} can be written discretely as,
\begin{equation}
\begin{aligned}
    p_i
    &= -\frac 1 {2(D-1)\pi}\sum_{j=1}^{N_v} \left[(\boldsymbol{\nabla} p)_j \cdot \frac{\boldsymbol{r}_{ij}}{r^D_{ij}}\right] \delta V_j
     + \frac 1 {2(D-1)\pi}\sum_{k=1}^{N_b} p_{b,k}\left(\frac{\boldsymbol{r}_{ik}}{r^D_{ik}} \cdot \delta \boldsymbol{S}_k\right) \\
    &= \sum_{j=1}^{N_v}\underbrace{-\frac {\delta V_j}{2(D-1)\pi} \frac{\boldsymbol{r}_{ij}}{r^D_{ij}} }_{ \boldsymbol{\displaystyle\alpha}_{ij}} \cdot \;(\boldsymbol{\nabla} p)_j +\sum_{k=1}^{N_b}\underbrace{ \frac 1 {2(D-1)\pi}\left(\frac{\boldsymbol{r}_{ik}}{r^D_{ik}} \cdot \delta \boldsymbol{S}_k\right)}_{\displaystyle \beta_{ik}}p_{b,k}  , 
     \label{eqn:GFI_dis}
\end{aligned}
\end{equation}
where $\boldsymbol{r}_{ij} = \boldsymbol{x}_j - \boldsymbol{x}_i$, $\boldsymbol{r}_{ik} = \boldsymbol{x}_k - \boldsymbol{x}_i$ are the displacement vectors from inner nodal points $\boldsymbol{x}_i$ to cell centers $\boldsymbol{x}_j$ and boundary element centers $\boldsymbol{x}_k$, respectively. the magnitude of the displacement vector is denoted as $r = |\boldsymbol{r}|$.
The compatibility condition on the boundary can be discretized as,
\begin{equation}
\begin{aligned}
    p_{b,k} & =-\frac 2 {2(D-1)\pi}\sum_{j=1}^{N_v} \left[(\boldsymbol{\nabla} p)_j \cdot \frac{\boldsymbol{r}_{kj}}{r^D_{kj}}\right] \delta V_j + \frac 2 {2(D-1)\pi}\sum_{k^\prime=1}^{N_b} p_{b,k^\prime}\left(\frac{\boldsymbol{r}_{kk^\prime}}{r^D_{kk^\prime}} \cdot \delta \boldsymbol{S}_{k^\prime}\right)\\
    & =\sum_{j=1}^{N_v} \underbrace{-\frac {2\delta V_j} {2(D-1)\pi}\frac{\boldsymbol{r}_{kj}}{r^D_{kj}}}_{2\displaystyle\boldsymbol{\alpha}_{kj}}\cdot \; (\boldsymbol{\nabla} p)_j  + \sum_{k^\prime=1}^{N_b}\underbrace{\frac 2 {2(D-1)\pi}\left(\frac{\boldsymbol{r}_{kk^\prime}}{r^D_{kk^\prime}} \cdot \delta \boldsymbol{S}_{k^\prime}\right)}_{2\beta_{kk^\prime}}p_{b,k^\prime} \\
    & = 2\gamma_k +  \sum_{k^\prime=1}^{N_b} 2\beta_{kk^\prime}p_{b,k^\prime}.
     \label{eqn:GFI_bc_dis}
\end{aligned}
\end{equation}
Here $\boldsymbol{r}_{kk^\prime} = \boldsymbol{x}_k - \boldsymbol{x}_{k^\prime}$ is the displacement vector from boundary element center $\boldsymbol{x}_k$ to $\boldsymbol{x}_k^\prime$. 
In the vicinity of $\boldsymbol{x}_k$ on the boundary, the displacement vector is tangent to the boundary and perpendicular to the normal vector $\delta \boldsymbol{S}_{k}$. 
Therefore,
\begin{equation}
    \lim_{\displaystyle\boldsymbol{x}_{k^\prime} \rightarrow \boldsymbol{x}_k} \left(\frac{\boldsymbol{r}_{kk^\prime}}{r^D_{kk^\prime}} \cdot \delta \boldsymbol{S}_{k^\prime}\right) = 0.
\end{equation}
Numerically, this singularity can be resolved by adding a small number (e.g. on the order of $10^{-10}$) to the denominator $r_{kk^\prime}^D$.

The equation \ref{eqn:GFI_bc_dis} can also be written in matrix form, if we denote $\boldsymbol{P}_b$ as the column vector containing all $p_{b,k}, k = 1,2, \ldots N_b$, 
\begin{equation}
    \boldsymbol{P}_b = 2\boldsymbol{\Gamma} + 2\mathbf{B}_b\boldsymbol{P}_b,
\end{equation}
where $\boldsymbol{\Gamma}$ containing all the $\gamma_k$, and $\mathbf{B}_b$ is the matrix with elements of $\beta_{kk^\prime}$.
Therefore, the boundary pressure can be solved by
\begin{equation}
    \boldsymbol{P}_b = 2(\mathbf{I} - 2\mathbf{B}_b)^{-1}\boldsymbol{\Gamma}.
    \label{eqn:boundaryPressure_inversion}
\end{equation}
For large-scale problems with more than thousands of boundary points, direct matrix inversion is expensive both for memory and computational time.
To compensate for such difficulty, we have utilized the conjugate gradient (CG) method \citep{stiefel1952methods,greenbaum1997iterative,wang2019spatial} to solve the linear system iteratively, which is equivalent to solving the modified equation,
\begin{equation}
    (\mathbf{I} - 2\mathbf{B}_b^T)(\mathbf{I} - 2\mathbf{B}_b)\boldsymbol{P}_b = 2(\mathbf{I} - 2\mathbf{B}_b^T)\boldsymbol{\Gamma}.
\end{equation}
The CG solver requires the matrix-vector multiplication for the matrices $\mathbf{I} - 2\mathbf{B}$ and $\mathbf{I} - 2\mathbf{B}^T$ to be calculated on the fly, without the need to store the whole matrix. The transposed system is implemented in the manner of an adjoint operator \citep{wang_wang_zaki_2022,wang2019discrete} and therefore circumvents the unfeasible memory requirement when the number of boundary points is large.  

In principle, the above linear system for boundary pressure should be linearly dependent, given that a difference in the mean pressure still results in the same gradient field.
Therefore, it is recommended to replace one equation with the condition $\boldsymbol{I}^T\boldsymbol{P}_b = 0$, declaring the mean pressure on the boundary is zero.

To summarize the process using matrix form, equations \ref{eqn:GFI_dis} and \ref{eqn:boundaryPressure_inversion} can be combined, and
\begin{equation}
    \boldsymbol{P} = \mathbf{A} \boldsymbol{\nabla P} + \mathbf{B} \boldsymbol{P}_b = \mathbf{A} \boldsymbol{\nabla P} + 2\mathbf{B}(\mathbf{I} - 2\mathbf{B}_b)^{-1}\mathbf{A}_b \boldsymbol{\nabla P} =  \underbrace{\left(\mathbf{A} + 2\mathbf{B}(\mathbf{I} - 2\mathbf{B}_b)^{-1}\mathbf{A}_b \right)}_{\mathbf{K}}\boldsymbol{\nabla P},
    \label{eqn:inversionmatrix}
\end{equation}
where $\mathbf{A}$ and $\mathbf{A}_b$ are the matrices with elements $\alpha_{ij}$ and $\alpha_{kj}$ in equations \ref{eqn:GFI_dis} and \ref{eqn:GFI_bc_dis}, respectively.

\subsection{Interpretation of Omnidirectional Integration (ODI)}
\label{Sec:InterpretationODI}

The design of the ODI algorithm was based on the observation of the scalar potential properties according to the field theory, which states that pressure is a scalar potential and its gradient forms a conservative field. 
\begin{figure}
    \centering
    \includegraphics[width=0.9\textwidth]{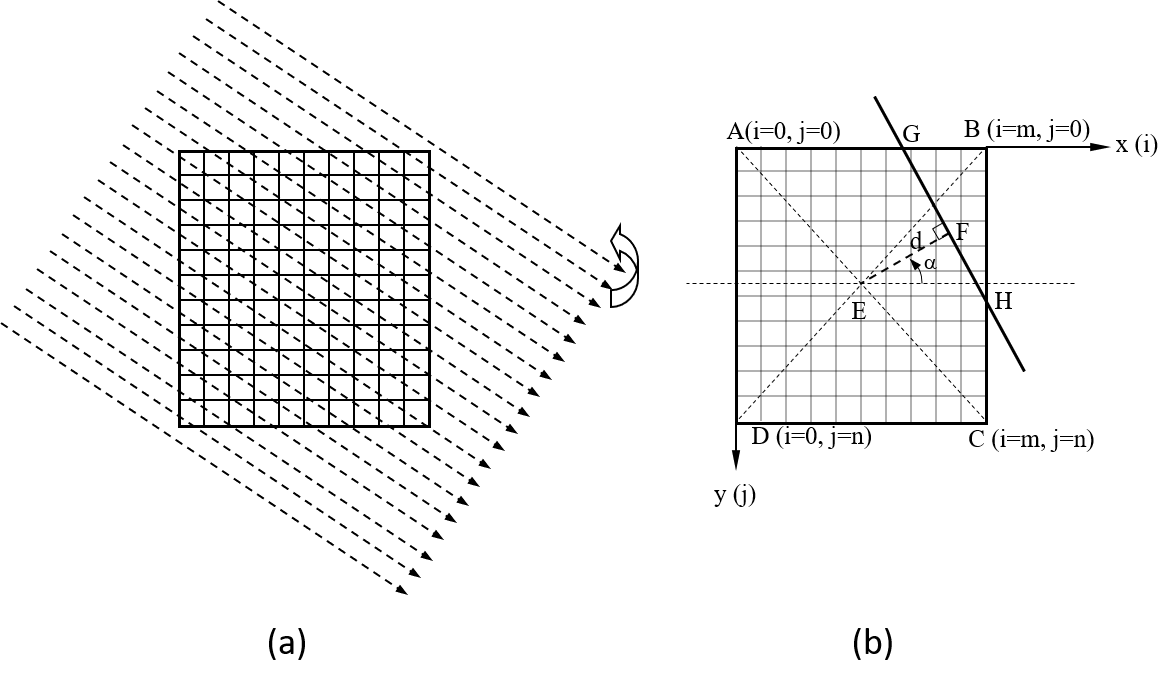}
    \caption{(a) The rotating parallel ray ODI algorithm. (b) Ray orientation with respect to the pressure calculation domain as a function of $d$ (distance from the domain center) and $\alpha$ (ray rotation angle).}
    \label{fig:path}
\end{figure}
Thus, spatial integration of the scalar potential gradient is independent of the integration path. With this understanding, the ODI algorithm first integrates the measured pressure gradient from all directions toward the point of interest, and then further averaging of the pressure values obtained along different integration paths allows the minimization of the effect of the errors embedded in the measured pressure gradient so as to obtain the final reconstructed pressure with improved accuracy. Specifically, the state-of-the-art parallel ray ODI method \citep{liu2016instantaneous, liu2020error} utilizes parallel rays as guidance for integration paths (see figure \ref{fig:path})  Effectively, by rotating the parallel rays, omnidirectional paths with equal weights coming from all directions toward the point of interest at any location within the computation domain can be generated. The orientation of the parallel rays is characterized by the angle $\alpha$ with respect to the horizontal direction of the pressure calculation domain. The distance between adjacent parallel rays is denoted as $\Delta d$, where $d$ is the distance from the domain center to the ray, i.e. the guideline for the pressure reconstruction path.
When $\Delta d$ and $\Delta \alpha$ become infinitesimal, the average of the line integrations becomes a two-dimensional surface integration with a convolution kernel that decays with $1/r$, where $r$ is the distance from the point of interest. This convolution kernel can be demonstrated to be equivalent to $\nabla G$ in an infinite domain where boundary conditions can be ignored. This sets the mathematical equivalence between ODI and GFI.

\begin{wrapfigure}{r}{0.3\textwidth}
    \centering
    \vspace{-3mm}
    \includegraphics[width = 0.18\textwidth]{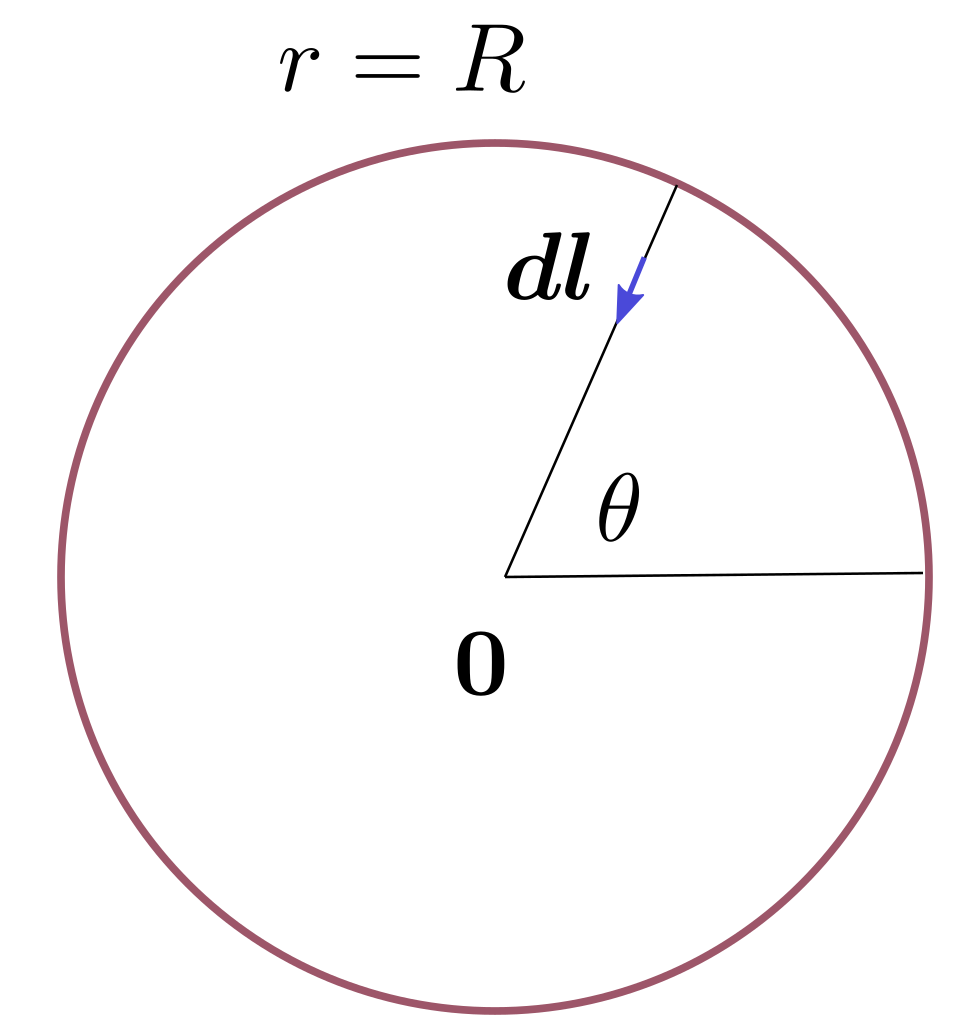}
    \caption{The integration path}
    \label{fig:path}
\end{wrapfigure}
We derive this equivalence in detail using parallel rays for a two-dimensional pressure field but is similar for a three-dimensional problem.
Suppose that the pressure gradient is given with noise, $\nabla p (x,y) = \nabla p_t(x,y) + \boldsymbol{\epsilon}(x,y)$, where $p_t$ is the true pressure field and $\boldsymbol{\epsilon}$ is the error or noise in the pressure gradient.
Also suppose the true pressure are given on the circle with radius $r=R$, namely $p(R\cos\theta,R\sin\theta)$ are given.
We pick a line at a angle $\theta$ with the x-axis, as shown in the schematic. 
For this single line integration, the reconstructed pressure at the origin is,
\begin{equation}
    \tilde{p}_{\theta}(\boldsymbol{0}) = p([R\cos\theta, R\sin\theta]) + \int_{r=R}^{r=0} \boldsymbol{\nabla} p \cdot \boldsymbol{dl} = p([R\cos\theta, R\sin\theta]) + \int_{r=0}^{r=R} \boldsymbol{\nabla} p \cdot \boldsymbol{e}_r dr
\end{equation}
If we use $M$ different radial integral lines, with gap in the angle, $\Delta \theta = \frac{2\pi}{M}$, the averaged pressure at the origin will be,
\begin{equation}
    \overline{\tilde{p}_{\theta}(\boldsymbol{0})}^\theta = \frac 1 M \sum_{i=1}^M p([R\cos\theta_i, R\sin\theta_i]) + \frac 1 M \sum_{i=1}^M \int_{r=0}^{r=R} \boldsymbol{\nabla} p \cdot \boldsymbol{e}_r dr, \quad \theta_i = \frac{2\pi i} {M}.
\end{equation}
Using the quantity $\Delta \theta = \frac{2\pi}{M}$, we can rewrite the above expression in integral form,
\begin{equation}
\begin{aligned}
    \overline{\tilde{p}_{\theta}(\boldsymbol{0})}^\theta &= \frac {1} {2\pi} \Delta \theta \sum_{i=1}^M p([R\cos\theta_i, R\sin\theta_i]) + \frac {1} {2\pi} \Delta \theta \sum_{i=1}^M \int_{r=0}^{r=R} \boldsymbol{\nabla} p \cdot \boldsymbol{e}_r dr, \quad \theta_i = \frac{2\pi i} {M}\\
    & = \frac {1} {2\pi} \int_{r=R} p d\theta  + \frac {1} {2\pi} \int_0^{2\pi} \int_{0}^{R} \boldsymbol{\nabla} p \cdot \boldsymbol{e}_r dr d\theta \\
    & = \frac {1} {2\pi} \int_{r=R} p d\theta  + \frac {1} {2\pi} \int_0^{2\pi} \int_{0}^{R} \boldsymbol{\nabla} p \cdot \frac{\boldsymbol{e}_r}{r} r dr d\theta\\
    & = \frac {1} {2\pi} \int_{r=R} p d\theta  + \frac {1} {2\pi} \iint_A \left(\boldsymbol{\nabla} p \cdot \frac{\boldsymbol{e}_r}{r}\right) dA.
    \end{aligned}
\end{equation}
Notice that the exactly equation holds for the true pressure field, namely,
\begin{equation}
    p_t(\boldsymbol{0}) = \frac {1} {2\pi} \int_{r=R} p_t d\theta  + \frac {1} {2\pi} \iint_A \left(\boldsymbol{\nabla} p_t \cdot \frac{\boldsymbol{e}_r}{r}\right) dA.
\end{equation}
Suppose that we have perfect boundary condition for pressure, e.g. $p = p_t$ at $r=R$, and that R is large enough.
The subtraction between the above two equations lead to,
\begin{equation}
    \overline{\delta p(\boldsymbol{0})} = \frac {1} {2\pi} \iint_A \left(\boldsymbol{\epsilon} \cdot \frac{\boldsymbol{e}_r}{r}\right) dA= \iint_A \left(\boldsymbol{\epsilon} \cdot \boldsymbol{\nabla} G(r)\right) dA.
\end{equation}

Therefore, the noise-reduction mechanism for ODI is a result of the integral kernel $\boldsymbol{\nabla} G(r)$.

\section{Results}
\label{Sec:results}
\subsection{Influence of pressure gradient perturbation at a single point}
\label{Sec:impulse}
As demonstrated in \S\ref{Sec:InterpretationODI}, the ODI method is equivalent to the GFI framework as they have exactly the same convolution kernel $\boldsymbol{\nabla} G$ in an infinite domain. 
The convolution kernel $\boldsymbol{\nabla} G$ also represents the influence of a single-point perturbation in the pressure gradient onto the resulting reconstructed pressure field.

\begin{figure}
    \centering
    \includegraphics[width=\textwidth]{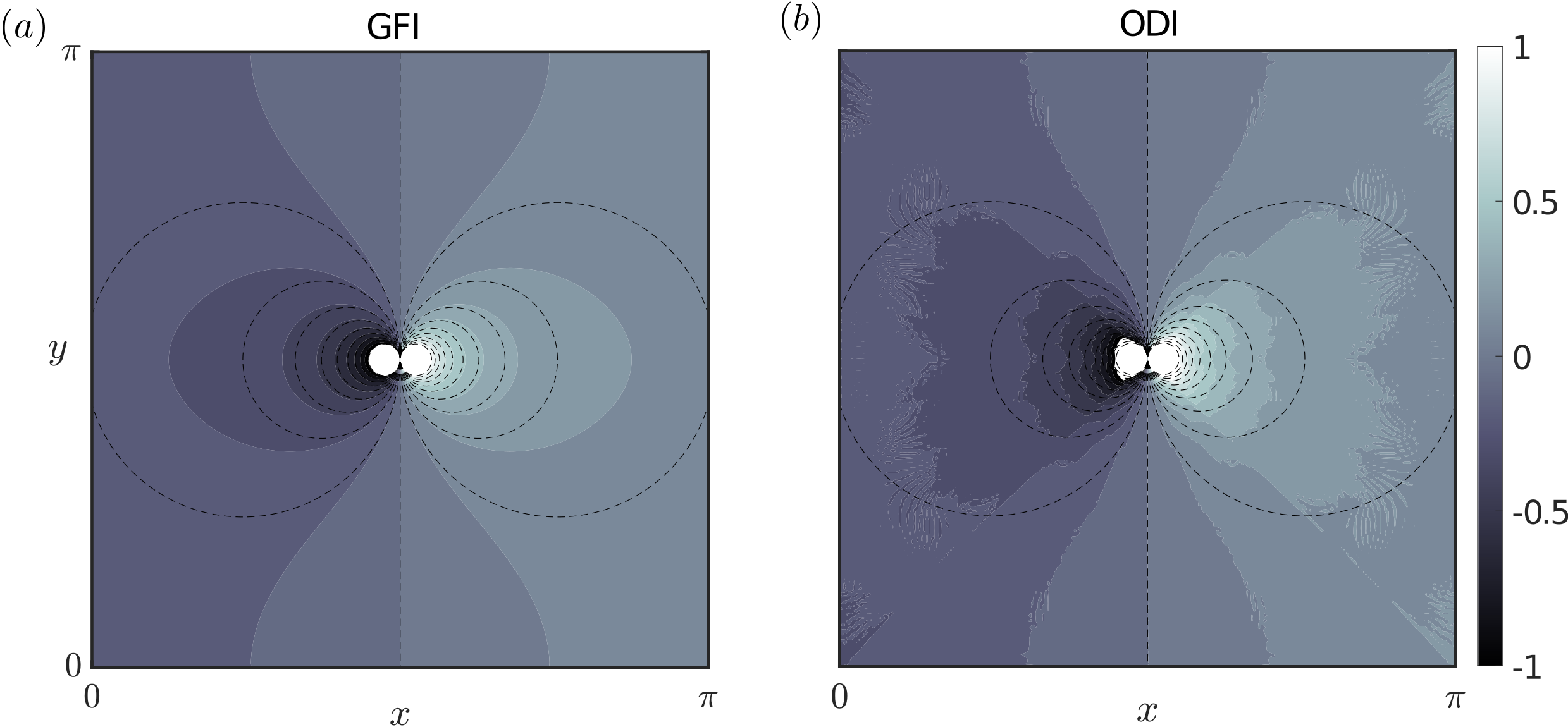}
    \caption{Comparison of the influence of localized perturbation in $\partial p /\partial x$ at the center of a square domain with the size ($\pi \times \pi$). The difference in resulting pressure fields is shown for GFI (left) and ODI (right). Dashed lines show the iso-contour of the horizontal component of $\boldsymbol{\nabla} G$ with the same contour levels.}
    \label{fig:DOD}
\end{figure}

Therefore, as a solid validation, evidence of the equivalence between ODI and GFI is shown in figure \ref{fig:DOD}, where the perturbation in the reconstructed pressure field is shown as a result of perturbation in the pressure gradient at the center of a finite, square domain of size $\pi \times \pi$.
We pick $\boldsymbol{\epsilon} = [\exp \left[-\frac{1}{2\pi \sigma^2} \left((x-\pi/2)^2 + (y-\pi/2)^2\right) \right] ,0]$ as a concentrated perturbation on the pressure gradient field, where $\sigma = 0.01$ is the width of the Gaussian. The difference in the reconstructed pressure fields using GFI and ODI are shown as colored contours in the left and right panels.
Dashed lines mark the iso-lines of the streamwise component of the convolution kernel $\boldsymbol{\nabla} G$. 
Despite that the ODI method exhibits spurious contour lines because of the zigzag integration paths, it is apparent that GFI and ODI have very similar behavior, with the pressure perturbation agreeing with $\frac{\partial G}{\partial x}$ near the center of the domain.
In fact, in an infinite domain, it is anticipated that $\delta p$ agrees perfectly with $\boldsymbol{\nabla} G$.
However, in an infinite domain, the boundary condition affects the pressure perturbation field $\delta p$, causing the pressure perturbation $\delta p$ to deviate from $\boldsymbol{\nabla} G$.

\subsection{Reconstruction of turbulent pressure field in a two-dimensional plane}
\label{Sec:2D}
We first apply Green's function integral method in a canonical setup of forced isotropic turbulence.
Pressure gradient field $\nabla p$  on a Cartesian grid of size $254 \times 254$ within a square domain of size $\frac{\pi}{2} \times \frac{\pi} 2$ are obtained from the Johns Hopkins Turbulence data based (JHTDB) \citep{perlman2007data,bappy2019lagrangian}.
It should be noted here that although the turbulent field is three-dimensional, the pressure reconstruction on a plane can be done using only two components of the pressure gradient.
For the reconstruction, uniform-distributed, independent noise of $0.4 |\boldsymbol{\nabla} p|_{max}$ is added to the pressure gradient at every observation.  
The true pressure is obtained at the inner nodal points and the boundary centers for comparison. This dataset has been used previously to demonstrate the accuracy of pressure reconstruction in case of realistic noisy measurements \citep{liu2020error}.
\begin{figure}
    \centering
    \includegraphics[width = \textwidth]{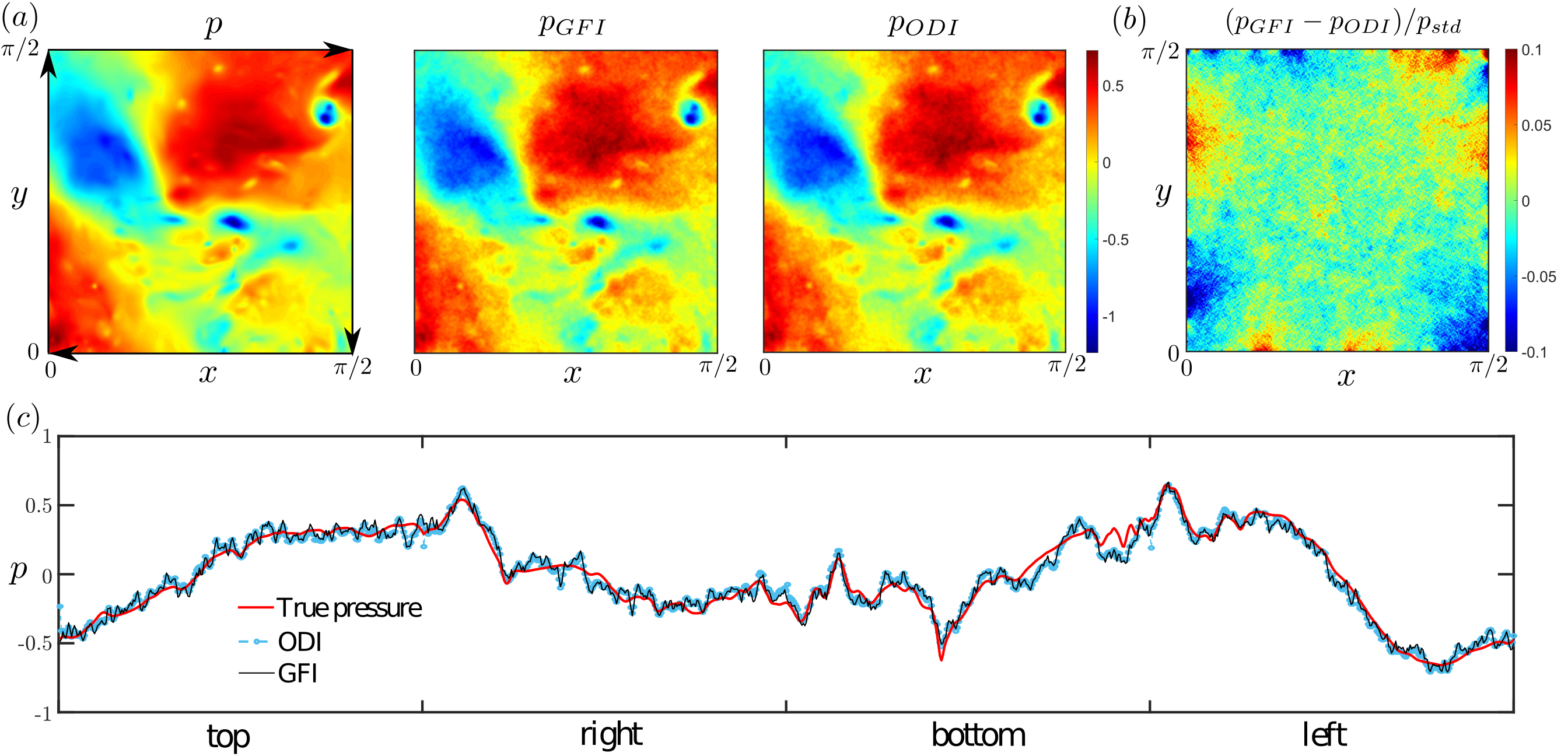}
    \caption{(a) Contour plots of the true pressure and reconstructed pressure from GFI and ODI. The reconstructions utilized error-embedded observations of pressure gradient. (b) the relative error between the reconstructed pressure using GFI and ODI methods. (c) Comparison of true and reconstructed pressure distribution over the boundary of the square domain.}
    \label{fig:GFI_ODI_compare_sample1}
\end{figure}
Comparisons between the true and sample reconstructed pressure fields using both GFI and ODI are shown in figure \ref{fig:GFI_ODI_compare_sample1}. The GFI and ODI methods yield very similar results, with a relative difference within 10\% of the standard deviation of the isotropic turbulence pressure fluctuation (i.e., a relatively small value), as shown in panels (a) and (b). The reconstructed boundary pressure shown in panel (c) has additional fluctuations due to the added noise in the pressure gradient. Nevertheless, the agreement with the true pressure is reasonable. 

\begin{table}
\begin{tabular}{ccccccc}
 \hline\\
Method                                                 & Grid size  & Noise level & $\displaystyle\overline{\frac{\epsilon_{p,std}}{p_{std}}}$ & $\displaystyle\left(\frac{\epsilon_{p,std}}{p_{std}}\right)_{std}$ & $\displaystyle\left(\frac{\epsilon_{p,std}}{p_{std}}\right)_{max}$ & CPU time(s)\\ \\\hline
ODI & $254 \times 254$ & 40\%        & 0.149                                       & 0.0153                                             & 0.23          & 59.6                                     \\
GFI                           & $254 \times 254$ & 40\%        & 0.143                                       & 0.0156                                             & 0.23 & 4.3 \\ \hline
\end{tabular}
\caption{Comparison between GFI and ODI methods.}
\label{TABLE:stats}
\end{table}
Statistics demonstrating the accuracy of ODI and GFI are shown in table \ref{TABLE:stats}. Taking a testing approach similar to \cite{liu2020error}, we generated 1000 samples of error-embedded observations of pressure gradient and summarized the mean, standard deviation, and maximum error in the reconstructed pressure.
A good agreement is demonstrated between ODI and GFI methods. In addition, the GFI requires much less computational time and is 14 times faster than ODI for the case tested, due to the exemption from zigzag integration paths.

\subsection{Denoising effect and spectral analysis}
\label{Sec:2DUQ}
From the original expression of pressure reconstruction \ref{eqn:GFI}, the convolution kernel $\boldsymbol{\nabla} G$ is responsible for the denoising effect of ODI and GFI.
In a finite domain where we can ignore the boundary conditions, the effect of applying the convolution is equivalent to modifying the Fourier components of the pressure gradient fields. However, in a finite domain, the boundary effect cannot be ignored and the eigenmodes of the operator in \ref{eqn:inversionmatrix} are not all Fourier modes.

\begin{figure}
    \centering
    \includegraphics[width=0.9\textwidth]{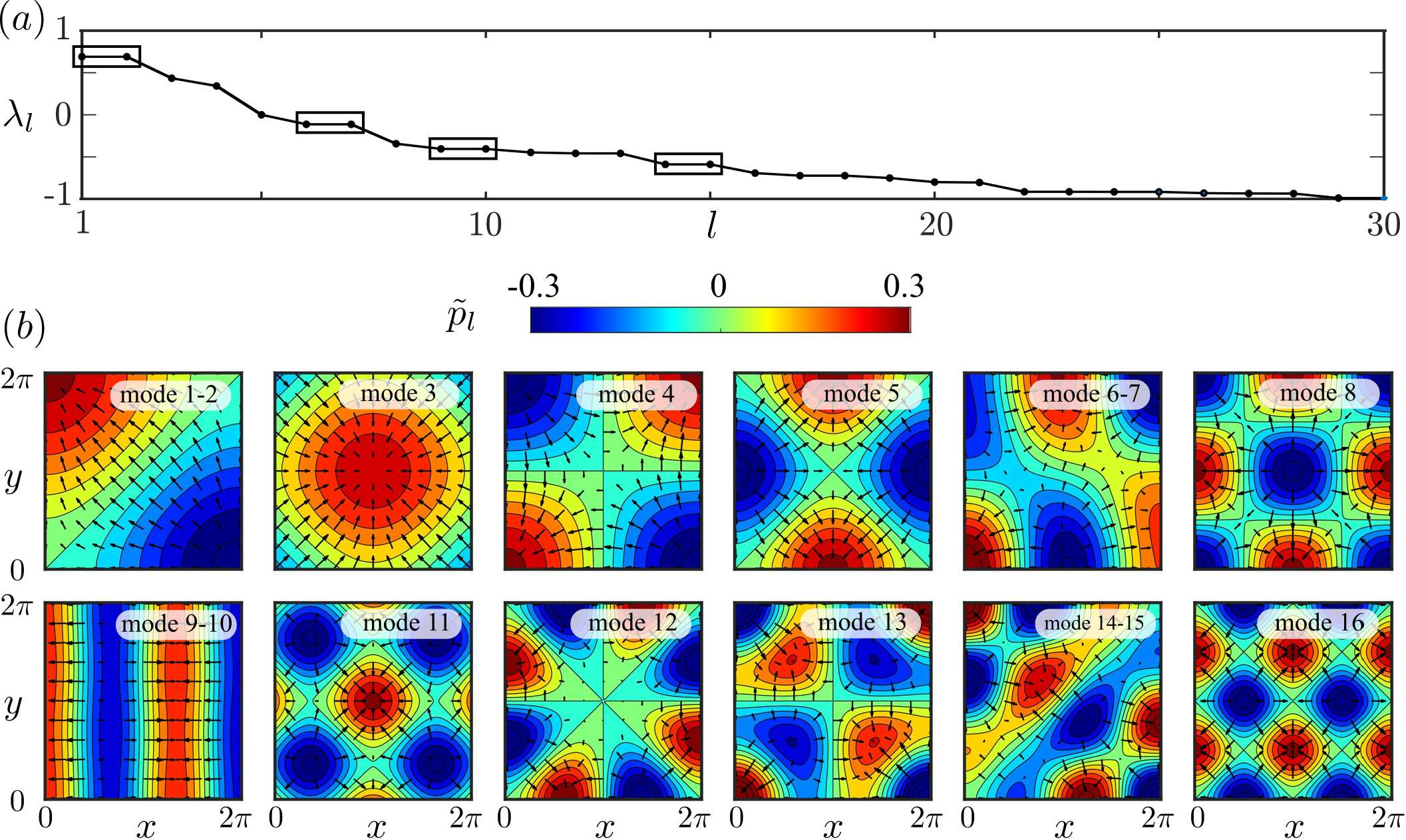}
    \caption{(a) Singular values for the leading 30 modes of matrix $\mathbf{K}$ in equation \ref{eqn:inversionmatrix}. (b) The first 16 singular vectors of matrix $\mathbf{K}$, the left and right singular vectors, $\tilde{p}_l$ and $\left(\boldsymbol{\nabla} \tilde{p}\right)_l$, are shown in colored contours and vectors, respectively. Pairs of eigenmodes, as marked in (a), are shown only once.}
    \label{fig:2deigAnalysis}
\end{figure}

For two-dimensional pressure reconstruction in a square domain, we computed the singular values and vectors of matrix $\mathbf{K}$ in equation \ref{eqn:inversionmatrix}, namely
\begin{equation}
    \mathbf{K} = \sum_{l=1}^{N_i} \lambda_l \tilde{p}_l \left(\boldsymbol{\nabla} \tilde{p}\right)^T_l,
\end{equation}
where $\tilde{p}_l$, $\left(\boldsymbol{\nabla} \tilde{p}\right)_l$, $\lambda_l$ are the left and right singular vectors and singular values, respectively. 
Through the Singular Value Decomposition of $\mathbf{K}$, we achieved modal analysis for the relationship between pressure gradient and reconstructed pressure field in the square domain. The leading singular values and vectors are shown in figure \ref{fig:2deigAnalysis}.
A majority of the modes are Fourier because of the homogeneous convolution kernel in a regularized domain. Nevertheless, the effect of a finite domain is apparent and creates radial modes (e.g. modes 13, 14).

This spectral analysis is helpful to understand the accuracy of GFI when observations of the pressure gradient are noisy. Suppose that the noise in the pressure gradient at each observation location has identical, independent Gaussian distribution with standard deviation $\tilde{\sigma}_{\boldsymbol{\nabla} p}$, the linear transformation $\mathbf{K}$ transfers the noise into the reconstructed pressure, with the covariance matrix,
\begin{equation}
    \mathcal{C}(\boldsymbol{\epsilon}_p) = \mathbf{K} \left(\tilde{\sigma}_{\boldsymbol{\nabla} p}^2 \mathbf{I}_{2N_v}\right) \mathbf{K}^T = \tilde{\sigma}_{\boldsymbol{\nabla} p}^2 \mathbf{K}\mathbf{K}^T.
\end{equation}
The number $2N_v$ is due to the two components of the pressure gradient in a two-dimensional setup. The standard deviation of the error in the pressure, is thusly
\begin{equation}
    \tilde{\sigma}_p = \sqrt{\frac{1}{N_i}Tr(\mathcal{C}(\boldsymbol{\epsilon}_p))} = \tilde{\sigma}_{\boldsymbol{\nabla} p} \sqrt{\frac{1}{N_i}Tr\left(\mathbf{K}\mathbf{K}^T\right)} = \tilde{\sigma}_{\boldsymbol{\nabla} p} \sqrt{\frac{1}{N_i}\sum_{l=1}^{N_i} \lambda_l^2}.
\end{equation}

\begin{figure}
    \centering
    \includegraphics[width=\textwidth]{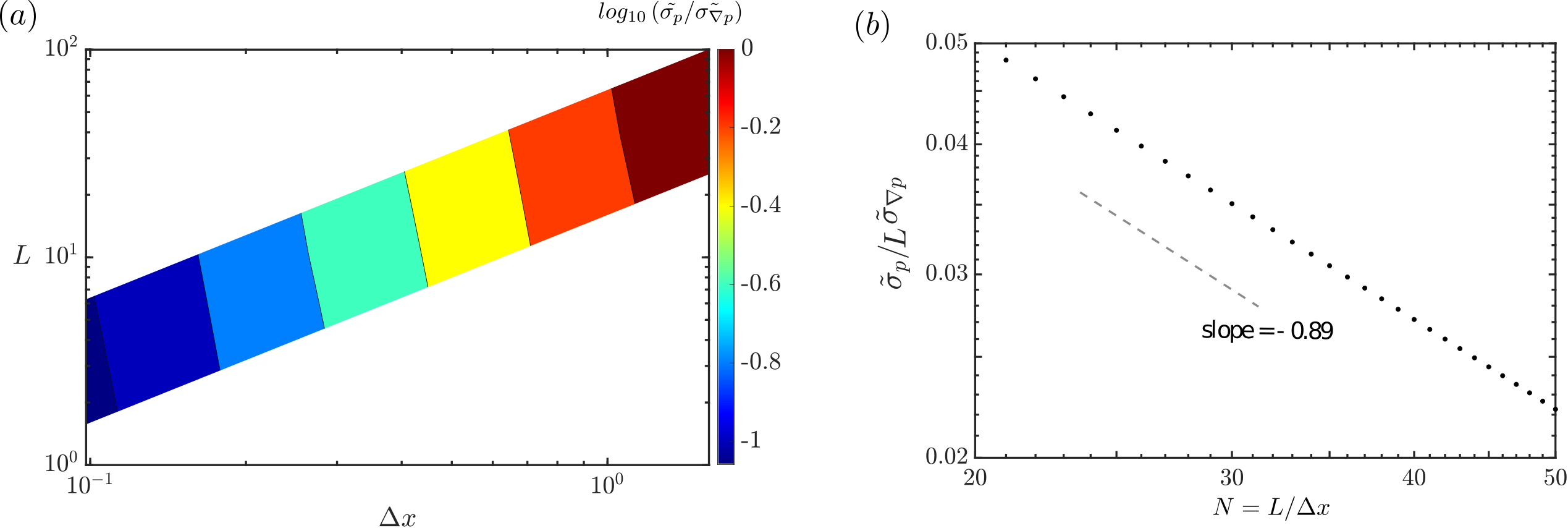}
    \caption{(a) The logarithmic of error attenuation factor $\tilde{\sigma}_p/\tilde{\sigma}_{\boldsymbol{\nabla} p}$ versus the grid spacing $\Delta x$ and $L$. (b) Demonstrating of the power law, $\tilde{\sigma}_p/(L\tilde{\sigma}_{\boldsymbol{\nabla} p})= aN^{-\kappa}$.}
    \label{fig:2DUQ}
\end{figure}

Therefore, the noise in the reconstructed pressure is attenuated with a factor $\tilde{\sigma}_p/\tilde{\sigma}_{\boldsymbol{\nabla} p}=\sqrt{\frac{1}{N_i}\sum_{l=1}^{N_i} \lambda_l^2}$. We focus on the square domain with uniform Cartesian grids and evaluate this factor from the eigenvalues of $\mathbf{KK}^T$ for different domain lengths $L$ and grid spacing $\Delta x$. 
The results are plotted in figure \ref{fig:2DUQ}(a). The log-log plot shows an apparent trend of a non-dimensionalized potential power law of the form $\tilde{\sigma}_p/(L \tilde{\sigma}_{\boldsymbol{\nabla} p})= a N^{-\kappa}$.
Using least-square fitting, we find that $\kappa = 0.89$ in the current case, and $a = 0.7239$, the result showing this power law is plotted in \ref{fig:2DUQ}(b).

\subsection{Three-dimensional pressure reconstruction within a multi-connected domain}
\label{Sec:3D}
The GFI framework we present here is general and can be applied to three-dimensional problems with complex geometry, including multiply connected domain \citep{gluzman2017computational}, i.e., domain with void regions inside. To demonstrate the capability of the GFI algorithm, we pick a challenging case of reconstructing pressure with observations of pressure gradients randomly distributed inside a unit cube. 
In addition, a spherical void region without any observations is allocated at the center of the cube.
This setup is relevant to scenarios where we have PIV data for a rising bubble in liquid, for example.

\begin{figure}
    \centering
    \includegraphics[width=\textwidth]{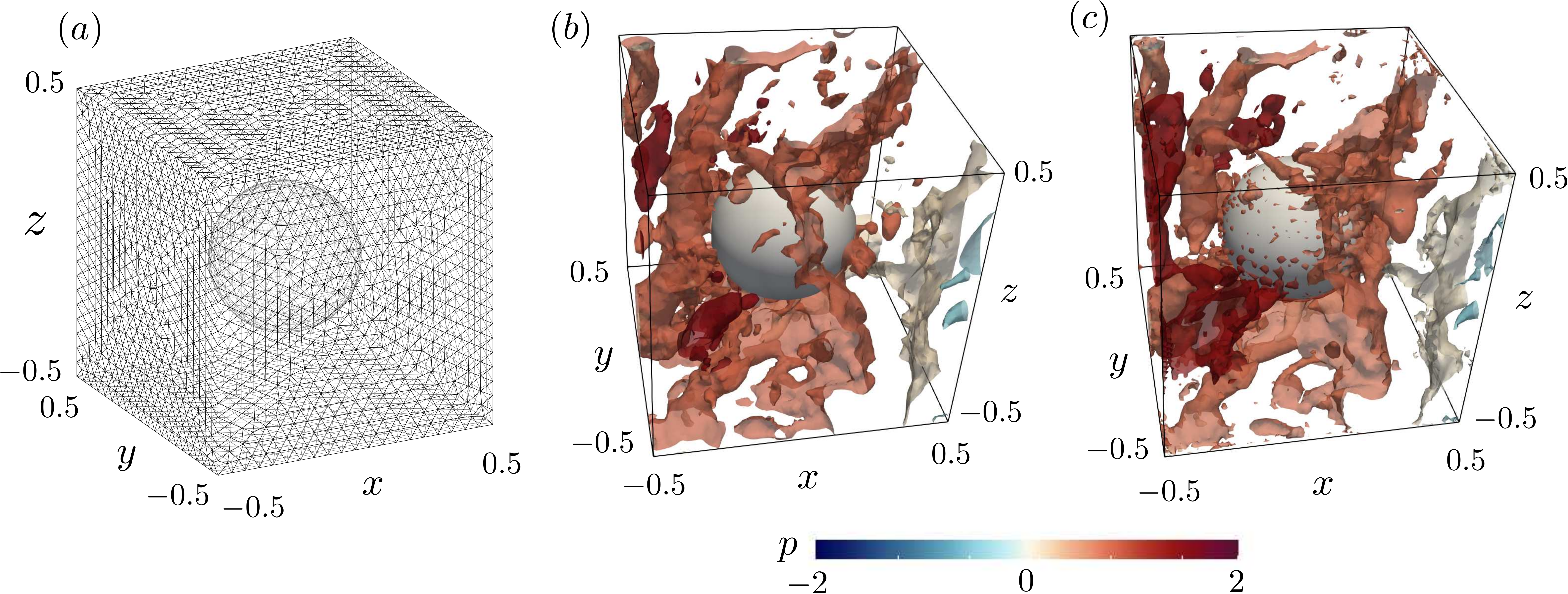}
    \caption{\del{Singular values vectors for the leading 30 modes of the pressure reconstruction.} Sample pressure reconstruction using GFI in a three-dimensional domain. (a) 3D domain and grids for pressure reconstruction; (b) Iso contour of true pressure field; (c) Iso-contour of the reconstructed pressure field.  }
    \label{fig:3Drecon}
\end{figure}

For this three-dimensional pressure reconstruction, a tetrahedral unstructured mesh is constructed inside the multi-connected domain, as shown in figure \ref{fig:3Drecon}(a).
Pressure gradients are obtained at the cell centers of the mesh from the turbulent database, and the pressure field is reconstructed at the nodal points.
Figure \ref{fig:3Drecon}(b) and (c) show iso-contours of the true and reconstructed pressure fields. The relative difference is 2\%. However, the current version of the algorithm does not involve higher-order interpolations, and the tetrahedral mesh creates extra spurious behavior in the reconstructed pressure field.

\begin{figure}
    \centering
    \includegraphics[width=\textwidth]{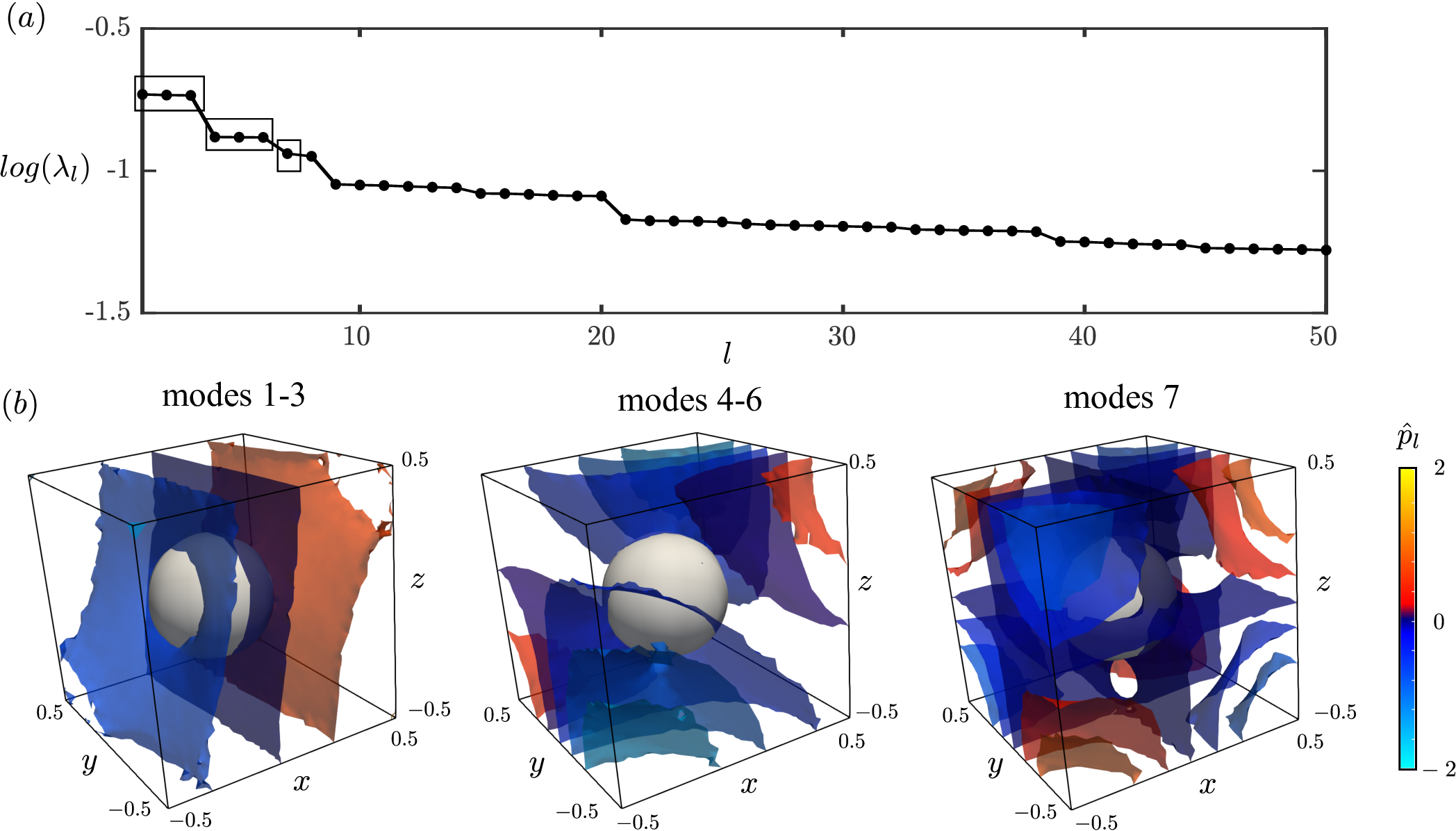}
    \caption{(a) Singular values for the leading 50 modes of the three-dimensional pressure reconstruction. (b) Leading left singular vectors of matrix $\mathbf{K}$, modes appearing in triads are only plotted once.}
    \label{fig:3D_eig}
\end{figure}

Furthermore, SVD is performed for matrix $\mathbf{K}$ for this three-dimensional reconstruction. The leading 7 modes are plotted in figure \ref{fig:3D_eig}. The singular values of the 3-D case is smaller than the two-dimensional case in \S\ref{Sec:2DUQ}, meaning a stronger denoising effect in three dimensional scenario.
The iso-contours of the left singular vectors $\tilde{p}_l$ are shown in panel (b). Similar to two-dimensional cases, where singular modes appear in pairs, many singular modes in the 3-D case appear in triads, representing the same mode with three different orientations.

\section{Conclusion and future work}
In summary, Green's function integral (GFI) method offers a promising approach for pressure field reconstruction in fluid mechanics applications. The GFI method is shown to have a deep mathematical connection with the state-of-the-art omnidirectional integration (ODI) method, which is used for pressure reconstruction from measured pressure gradient fields. Comparisons of GFI and ODI show that they are equivalent in pressure reconstruction, while GFI appears to be more computationally efficient for both 2D and 3D domains, making it a suitable alternative for practical applications. In addition, the GFI method offers a way to quantify the denoising effect through eigenanalysis, with the reconstructed error decreasing as the number of samples increases, which is an important consideration for improving measurement accuracy. Future research may focus on improving computational efficiency further by exploring multi-grid formulations and applying the method to a wider range of geometries. Overall, the GFI method offers the potential for enhancing pressure measurement accuracy and efficiency in fluid mechanics research.


\bibliography{Manuscript}

\end{document}